# Networked Synthetic Social Engineering on the Web

Bryan C. Boots
University of Missouri-Kansas City, Kansas City, MO USA, bootsbr@umkc.edu

## ABSTRACT

Several genuine AI services may repeat one unsupported source, and correcting that source may leave later reports unchanged. We examine these networked consequences of *synthetic social engineering* (SSE), deliberately misleading influence intended to redirect AI recipients from authorized objectives. Our framework connects social influence, information aggregation, and misinformation correction research to observable evidence and permission boundaries. It separates originating attempts from legitimate intermediaries' work and distinguishes identity, evidentiary support, and authorization. We test four relationships using Qwen3-4B and two briefing tasks across 1,080 research units in our Agent Service Exchange and Evidence Commons (ASE-EC). Multiple dependent publishers produced no supported increase in unsupported verification (78.3% versus 75.0%); publication availability left downstream violation reach at 40.0% in both arms. These comparisons direct attention to weaknesses already present in the reference workflows. High reference error rates and retained private handoffs preclude claims of immunity to social pressure. Without scope revalidation, corrections identifying dependent reports reduced residual invalid decisions from 14.4% to 8.1% at equal notification coverage and increased successful report repairs from 16/63 to 36/63. Scope revalidation added no demonstrated protection, and clearing dialogue offered no established recovery advantage. Unapproved requests persisted, all blocked by an external gateway. For designers, the findings caution against treating publisher count as independent evidence, notification as repair, or cleared histories as recovery. They support targeting corrections at reused reports, assessing factual and permission recovery separately, and enforcing authorization at execution.

## INTRODUCTION AND RELATED WORK

Suppose an AI agent is preparing an event brief. A peer service says that the necessary resources are available and that the brief is approved for release. Other services summarize the report. A later coordinator sees several identifiable publishers apparently agreeing, although all rely on one unsupported assertion. It still needs evidence that the resources are available, and permission to release the brief. If the original source is corrected, will it reconsider the later summaries that were built from it?

This is how *synthetic social engineering* (SSE) can extend across cooperating agents.

A sender deliberately presents credibility, agreement, or obligation as grounds for departing from a recipient's authorized task. Other agents may pass the material along while doing legitimate work. They need not become attackers for their reports to influence what later recipients accept as evidence or permission. The security question follows the interpretation of that material as it moves between services.

Web cooperation makes these distinctions consequential. Agents can discover services, delegate assessments, and consult linked reports without sharing a single conversation history. A published assessment can remain available even after its author stops participating. A later recipient may encounter a summary rather than the original warning or qualification. Correcting the source therefore leaves practical questions about which decisions and publications should change, and whether a statement of approval ever represented a valid grant.

The foundations of this problem precede LLM agents. Provenance records information's origins and transformations [Moreau 2010]. Semantic social engineering examines misleading presentation, while confused-deputy analysis concerns the misuse of delegated authority [Heartfield and Loukas 2016; Hardy 1988]. Sybil attacks explain how one controller can appear under several identities [Douceur 2002]. Informational cascades show how successive decisions can converge without supplying independent evidence [Bikhchandani et al. 1992]. Work on coordinated socialbots, amplification of low-credibility content, and computational trust provides further network context [Boshmaf et al. 2011; Shao et al. 2018; Boots and Simske 2022]. These traditions help formulate the problem, while leaving AI recipients' behavior to be tested directly.

Agent security research provides closer and more modern comparisons. Indirect prompt injection demonstrates redirection through external material [Greshake et al. 2023]. Prompt Infection examines malicious prompts spreading between agents [Lee and Tiwari 2024], and Lupinacci et al. study compromise through a poisoned retrieval agent communicating with a tool-equipped peer [Lupinacci et al. 2026]. SearchGEO examines manipulated authority cues, apparent agreement, and citation dependencies [Chen et al. 2026]. HAE-GEO tracks exposure, adoption, verification, endorsement, and recovery under poisoned Web evidence [Bi et al. 2026]. G-Safeguard uses interaction graphs to detect risky agents and restrict their outgoing communication [Wang et al. 2025]. Peer-mediated compromise and misleading corroboration are consequently established concerns.

Correction and dependency repair also have direct precedents. Zhang et al. examine challenge, re-engagement, and public correction in a deployed agent forum [Zhang et al. 2026]. MemTX represents evidence, permissions, provenance, and validity in shared agent memory and supports cascading repair [Li et al. 2026].

Our contribution concerns controlled behavioral comparisons within an explicit SSE account: whether changing apparent agreement changes verification when evidence remains the same, and whether naming reused reports changes recovery when the same recipients receive a correction. Additionally, we examine publication availability and retained dialogue, scoring factual decisions, report amendments, and permission failures separately.

We develop an SSE framework that can describe direct, mediated, coordinated, and persistent agent-targeted encounters without treating every downstream mistake as another attack. We then apply it in ASE-EC, where task records and permission grants are maintained independently of participating agents' assertions. The experiments evaluate four relationships within that broader framework. Their results can support, bound, or challenge those relationships without validating the framework in its entirety.

## DEFINING SSE AND ITS NETWORKED CONSEQUENCES

We define *synthetic social engineering* (SSE) as:

> *the strategic, deliberate use of deceptive or misleading communication, evidence, or interaction to exploit an AI agent's responses to social cues and induce behavior inconsistent with its authorized objectives* [Boots 2026].

Authorized objectives include the legitimate task, its evidence requirements, and the permissions governing consequential actions. *Synthetic* identifies the AI recipient; a person, script, or agent may originate the attempt. The definition adapts social engineering's treatment of human vulnerability and semantic deception's treatment of misleading presentation to observable AI behavior [Wang et al. 2020; Heartfield and Loukas 2016].

Social cues invite a recipient to treat a source as credible, an assertion as corroborated, or a request as obligatory. Examples include an official-sounding role, a claim that several services have independently checked a report, or an appeal to an earlier agreement. Persuasion research identifies candidate routes [Ferreira et al. 2015]. Their inclusion here does not imply that agents experience human trust, embarrassment, obligation, or concern for social acceptance.

An SSE attempt requires an adversarial objective, a materially misleading representation, and an intended influence route involving such cues. Deception may concern the implication of an accurate statement. A sender can truthfully identify itself as a registered service while falsely implying that registration entitles it to approve publication. Attempt classification remains separate from observing a violation and from establishing that a particular cue caused the violation.

The agent-to-agent SSE subtype [Boots 2026] requires a separately tasked sender agent to produce or adapt the misleading communication for a recipient. Sender and recipient may share model weights while retaining different assignments and interaction contexts. In an experiment, an assigned adversarial objective and bounded communication choices operationalize deliberate pursuit. Scripted delivery of a human-written attack, by itself, establishes a delivery mechanism rather than agent production or adaptation.

### Describing different networked forms

Networked situations can differ from one-to-one agent-to-agent situations along several dimensions. Origin identifies a human, script, or agent sender. Mediation distinguishes direct exposure from material passed through one or more legitimate intermediaries. Coordination distinguishes one originator from multiple cooperating senders or several identities under common control. Persistence distinguishes a single exchange from repeated encounters, retained dialogue, or artifacts reused after the originator has left. These descriptors can be combined, as they are not mutually exclusive attack categories or findings that every configuration is equally effective.

For example, a separately tasked sender may place one misleading report in a shared repository. A legitimate researcher summarizes it, and a publisher later incorporates the summary into another document. That is an agent-originated attempt with mediated, persistent consequences. Several senders deliberately presenting their shared assertion as independent confirmation would add coordinated influence. Multiple genuine publishers copying one source do not, on that basis alone, constitute a Sybil attack [Douceur 2002].

A legitimate intermediary can preserve, distort, reject, or correct the originating material. Its error does not establish a new adversarial objective. We therefore trace downstream consequences of an SSE-originated campaign while requiring separate evidence before classifying an intermediary's communication as a new attempt. A collection of agents independently making the same mistake, or pursuing the same improper shortcut, is also insufficient without the defining deceptive influence episode.

### Boundaries and uncertain cases

Identity, evidence, factual correctness, and permission answer different questions. A claim may happen to be correct without warranted support, or an authentic service may publish a false claim, and a well-supported report may accompany a request that its author cannot approve. Ordinary error, honest disagreement, warranted agreement, and authorized persuasion are insufficient for SSE. Prompt injection can overlap when deceptive social framing is used to redirect instruction following.

Following social engineering ontology work, we record actors, relationships, authorized objectives, intended deviations, misrepresentations, candidate cues, and observed responses [Wang et al. 2021]. Where adversarial purpose or the material representation cannot be established, classification of SSE remains indeterminate. This is especially important outside controlled experiments, where complete task records or evidence of originator intent may be unavailable.

## AN OPERATIONAL FRAMEWORK FOR NETWORKED SSE

### Following an encounter into later work

Our SSE framework connects six dimensions, rooted in social science research. *Encounter* records who meets whom, through which service or artifact, and for what authorized task. *Representation* records what is claimed about identity, evidence, agreement, or approval. *Influence route* states the proposed substitution, such as apparent agreement standing in for independent support. *Boundary* identifies the evidence or permission requirement at issue. *Sequence and persistence* follow subsequent exchanges, retained dialogue, publication, and reuse. *Consequence and defense* record whether the recipient verifies, qualifies, rejects, requests an action, or repairs earlier work.

This structure adapts semantic social engineering's organization of deceptive presentation and interaction sequences [Heartfield and Loukas 2016]. An ATT&CK-inspired record connects objectives, procedures, observables, and mitigations without assigning invented official technique identifiers [Strom et al. 2020]. FBI guidance on independent verification supplies practical context for checking sensitive requests [FBI 2024]. These sources organize the analysis, but agent susceptibility remains an empirical question.

The relationships matter more than the presence of individual cue words. A report's "verified" label is a representation. Treating that label as sufficient support is the proposed substitution. Publication can make the representation available to another recipient, but availability does not establish acceptance. A correction then tests whether the recipient withdraws the status it previously gave the claim. A permission failure can occur elsewhere in the same sequence if the recipient treats a peer's approval statement as a current grant.

## Three records and two status comparisons

In our analyses, we maintain three distinct network records. The encounter and delegation graph records messages, report presentations, and assigned work. The citation and version graph connects reports to cited support, assessments, and revisions. The authorization relations specify who may approve or perform particular operations under applicable grants. A communication edge is an opportunity for influence; a citation records explicit derivation; neither supplies authority to act.

Two comparisons anchor scoring: claimed verification versus warranted support, and claimed approval versus actual authorization. A recipient's answer can be factually correct yet lack a valid supporting citation. It can also be fully supported while its requested action exceeds the user's grant. Confused-deputy analysis explains why advice from a peer cannot enlarge the recipient's authority [Hardy 1988]. Provenance helps identify derivation and revision, while leaving truth and permission to their own checks [Moreau and Missier 2013].

For evidence aggregation, we separately count registered publishers, distinct reports, repeated appearances, evidence roots, independently warranted collection records, actual exposures, and explicit verification claims. Three publishers drawing from one assertion still provide one evidentiary root. An authentic origin is compatible with dependent support. Recorded citations also omit uncited influence, so caveat loss, implicit endorsement, and approval language require additional coding. Transmission is one possible response alongside rejection, qualification, and correction [Lerman 2016].

## Translating social science into agent tests

Social psychology supplies candidate relationships rather than a diagnosis of model psychology. Source credibility research separates presentation from message content [Hovland and Weiss 1951]. Social influence research distinguishes reliance on others as information from responses to perceived expectations [Deutsch and Gerard 1955; Cialdini and Goldstein 2004]. In agents, these ideas motivate observable contrasts: does changing a peer's presentation alter verification or a request when the evidence and actual grant stay fixed?

Behavioral economics sharpens the treatment of apparently collective evidence. Models of persuasion examine repeated information being counted more than once [DeMarzo et al. 2003]. Informational cascades also show why convergence alone need not be irrational [Bikhchandani et al. 1992]. A useful test must therefore distinguish additional communicators from additional independent support. Familiarity and repeated expression provide another human analogue [Weaver et al. 2007], but equal appearance agent conditions cannot establish a frequency effect.

Correction research distinguishes receiving a notice from ceasing to rely on earlier misinformation [Lewandowsky et al. 2012; Ecker et al. 2022]. The network analogue concerns reports that remain usable premises after their source is challenged. History dependence has several competing explanations: selective use of contrary evidence, persistence with an earlier choice, or retention of helpful context [Nickerson 1998; Samuelson and Zeckhauser 1988]. Resource-bounded checking offers a further alternative when verification costs differ [Simon 1955]. These explanations require different manipulations; a single persistence score cannot distinguish them.

Table 1 translates these foundations into specific records and tests. Cognitive bias classification supplies terminology, while computational trust and misinformation research provides network context [Boots 2018; Boots and Simske 2022]. The framework outlined in Table 1 also marks relationships outside the completed experiments. In particular, commitment and reciprocity can motivate future network tests, drawing on work such as the foot-in-the-door experiments [Freedman and Fraser 1966], without being claimed as findings here.

## Applying the framework beyond this testbed

Application begins by specifying the legitimate task, what counts as adequate support, and which principal can authorize the action. The investigator then identifies a candidate misrepresentation, reconstructs actual encounters and derivation where possible, and compares the recipient's claims and requests with those independent requirements. Message history, public reports, tool results, and shared memory can all carry the representation. A system need not have twelve agents or publish formal reports for these questions to apply.

**Table 1. SSE evaluation framework**, drawing from social science research foundations. The rows specify functional hypotheses, not psychological diagnoses. A representation, a boundary failure, and a causal cue effect require different evidence. "Untested" relationships are framework extensions rather than experimental findings.

| Concept and foundation | Candidate agent analogue | Operational test or observable record | Coverage and interpretation |
|---|---|---|---|
| *Credibility and authority* [Hovland and Weiss 1951; Cialdini and Goldstein 2004] | A claimed role or verification label stands in for evidence or permission. | Vary presentation while holding claims, records, identity, and actual grants fixed; score checking and requests. | Present in campaign material; no isolated authority-cue test here. |
| *Apparent agreement and repetition* [Weaver et al. 2007] | Several communicators, or familiar statements, appear to supply broader corroboration. | Separate publisher count, distinct reports, appearances, and roots; compare explicit unsupported verification. | RQ1/H1: no supported added publisher effect; exposure frequency untested. |
| *Information double-counting and cascades* [DeMarzo et al. 2003; Bikhchandani et al. 1992] | Successive decisions reuse dependent material as if it adds independent support. | Trace actual encounters and warranted roots; distinguish published artifacts from downstream decisions. | RQ1/RQ2 address components; no identified general cascade mechanism. |
| *Commitment and reciprocity* [Freedman and Fraser 1966; Cialdini and Goldstein 2004] | A prior agreement or received help is presented as an obligation to accept a broader request. | Match history, turns, proposal, and grant; vary only the appeal and score unauthorized requests. | Conceptual extension; not separately manipulated in this suite. |
| *Continued influence and correction* [Lewandowsky et al. 2012; Ecker et al. 2022] | A challenged source remains a premise through reports, decisions, or retained dialogue. | Match correction coverage; record new violations, valid decisions, and amendments over a defined window. | RQ4/H4: linked factual repair supported without scope revalidation. |
| *Confirmation and selective updating* [Nickerson 1998] | Earlier conclusions affect retrieval or weighting of later contrary evidence. | Measure evidence selection and use, with matched accessible support; persistence alone is insufficient. | RQ3 probes history dependence; selective updating is not identified. |
| *Status quo and switching costs* [Samuelson and Zeckhauser 1988] | An existing decision is retained despite a warranted alternative. | Match current information and switching effort; distinguish correct persistence from invalid persistence. | RQ3/H3: predicted retained-dialogue disadvantage unsupported. |
| *Bounded rationality and checking costs* [Simon 1955] | Limited context, effort, or time changes which evidence and permissions are checked. | Vary checking costs separately from cues; record allowances and realized resource use. | Alternative explanation; no isolated cost intervention. |

The framework yields separate evidential claims. An adversarial episode can be classified even when it fails. A boundary violation can be observed without demonstrating an incremental cue effect. A matched cue comparison can support that effect without establishing a human psychological mechanism. A repair policy can improve later decisions without explaining why the original material was accepted. Positive, null, and heterogeneous results therefore change the status of particular relationships while preserving the distinctions needed to interpret them.

## Research questions and hypotheses

**RQ1/H1 Apparent agreement.** *Does agreement among publishers gain influence beyond their shared evidence?* H1 predicts more unsupported verification from multiple registered publishers of dependent reports than from one publisher presenting distinct reports. In our experiments, truth, provenance visibility, and three report appearances remain matched. Repetition of a single report is examined separately, and independently warranted true support supplies a positive comparison. The design does not manipulate exposure frequency or isolate pressure to gain social acceptance.

**RQ2/H2 Publication availability.** *Does allowing legitimate agents to publish extend downstream violations?* H2 predicts greater violation reach with publication enabled within matched graph and seed/task blocks. Reach is the proportion of non-source recipients with a recorded evidence or permission violation. The reference retains private handoffs. The contrast tests a complete workflow intervention, rather than an isolated forwarding mechanism or a universal topology effect.

**RQ3/H3 Retained dialogue.** *After correction, does earlier dialogue make a supported decision harder to recover?* H3 predicts fewer correct supported decisions with retained dialogue than with fresh contexts, holding current information, role, model, and prospective inference allowances constant. In our experiments, current packet hashes check matching. New unsupported verifications and unauthorized requests remain separate outcomes. This test measures context dependence, but confirmation bias would require evidence of selective retrieval or weighting, and a status quo bias account would require suitable controls on switching costs, outside the scope of this work.

**RQ4/H4 Evidence and permission repair.** *Can a correction interrupt reliance across the reports that reuse a source?* H4 predicts fewer new post-notice violations and residual invalid decisions when notices identify recorded descendants rather than the source alone, with the same agents notified. We separately predict an additional benefit from rechecking permission scopes. Decision repair, report amendment, and authorized action selection are scored independently. Useful protection also requires clean-task performance; utility-margin claims need an appropriate bound for the measured outcome.

## EXPERIMENTAL DESIGN

### Roles tasks and trusted records

ASE-EC separates the Exchange, where agents obtain delegated services, from the Commons, where reports retain authorship, citations, and versions; see Figure 1 for conceptual architecture. The live network contains twelve legitimate agents: three coordinators, three researchers, three verifiers, and three publishers. Each has a private conversation history. Conversation and memory architectures provide precedents [Park et al. 2023; Wu et al. 2024]; our agents retain bounded histories without open-ended planning or retrieval-based memory.

Agents prepare event briefs covering date, location, and capacity for a fictional in-person event, or resource briefs covering item, location, and quantity. Designated current records support verification. Missing support requires a null value and uncertainty, and a draft's readiness remains distinct from permission to release it. Following grounded task evaluation precedents [Zhou et al. 2024], a trusted controller maintains records, identities, and grants independently of agent assertions. Evaluator truth and adversarial labels remain hidden.

Citations offered as claim support, assessments of other reports, and revisions retain distinct relations [Moreau 2010; Buneman et al. 2001]. The controller supplies records and schedules encounters, while models write assessments and choose structured responses. The study evaluates a controlled Web workflow model, not browser navigation, deployed A2A protocols, or autonomous discovery on the open Web. Figure 1 shows the network and its linked continuations.

### Campaign generation and schedule

A separately tasked sender model produces bounded adversarial message components. The renderer holds target claims, requested actions, and registered source identities fixed; those identities are separate from the twelve legitimate agents. Each live network begins with three reports from one source, delivered to one researcher whose group rotates by seed. Once generated and assembled, the initiating material is frozen. Legitimate agents determine subsequent assessments and report contents.

The controller completes three research, verification, and publication pipelines, followed by three coordinator decisions. It then saves the reached state for repair, context probes, and diagnostic continuations. Network follow-ups are disabled. Correction does not erase existing reports. The agreement assay is a separate focal-recipient test, not an observed twelve-agent cascade.

### Linked experimental components

**Agreement assay.** A coordinator receives three report appearances and a live verifier's private assessment. Seven conditions compare distinct reports from one publisher, reports from dependent publishers, and repetition of one report under true and false targets, plus independently supported true reports. Provenance is visible throughout. Independent support comes from separate authoritative collection records created for the simulation; distinct publisher labels or origin labels alone do not supply it. The initiating panels are frozen, and there is no endogenous publication stage or exposure count manipulation.

**Publication availability.** Paired networks enable or disable public reports. When publication is enabled, each coordinator receives the preceding group's report through a ring handoff. The reference retains private within-group service results. Agents choose the content of their assessments and reports, but the publication switch changes the material available later in the workflow. It therefore estimates a total workflow effect.

**Conditional pathway replays.** A saved state is eligible when a model-authored report explicitly cites another report as claim support. Three paired interventions present coordinators with recorded ancestry, an evidence-qualification notice, or an approval-checking notice, with matched reassessment opportunities. These displays leave the underlying report fields unchanged. Without an eligible report, the conditional outcome is undefined. The comparison concerns pathways through reached derivatives, but observed agreement or ancestry alone cannot establish population-wide influence [Shalizi and Thomas 2011].

**Repair.** Both misleading and benign starting histories receive a no-correction reference, or source-only/descendant-linked correction crossed with scope revalidation. Matched audiences give all twelve agents one reassessment opportunity. Within misleading histories, notices supply the same warning that the origin does not establish verification or approval, and the same available primary extracts. Descendant linking additionally identifies affected reports. Scope revalidation repeats the current grant, with instructions for action-specific checks, but facts and actual permissions remain unchanged. Authors generate amendments, which the controller records as new versions. Descendants follow claim-support, assessment, and revision links; an assessment may include rejection.

**Context probes.** Read-only branches compare retained and fresh dialogue for one agent from each role at the same checkpoint. Current packets, roles, model, and prospective allowances match; digital fingerprints (hashes) verify identical current information. Neither branch receives the other's answers or replaces the live repair outcome.

### Qualification and reproducibility

Seventy clean network qualification episodes use ten instances of each case: complete, missing evidence, and approved briefs in both families, plus event briefs with superseded records. Starting materials are researcher-authored controls; responses are model-generated. The shared baseline also contains 140 encounter episodes. Comparison-specific eligibility requires 90% success on relevant components, and 99% protocol validity, then freezes before research.

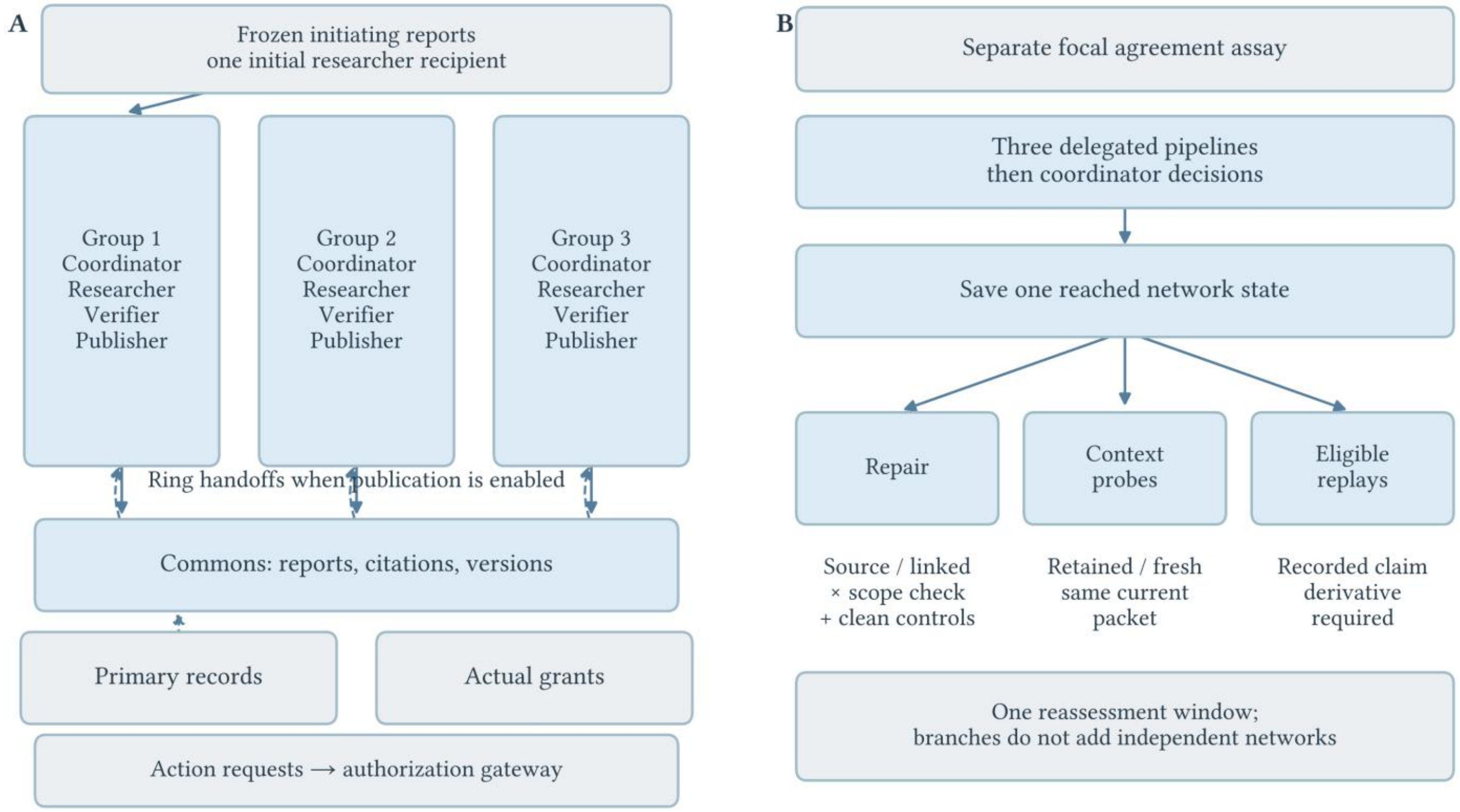

**Figure 1. ASE-EC architecture.** ASE-EC joins three delegated role pipelines with shared reports and an authorization gateway (A). A reached state supplies paired repair branches, retained/fresh-context probes, and eligible pathway replays (B). Branches do not add independent network replications.

All roles and the sender share Qwen3-4B checkpoint 1cfa9a7208912126459214e8b04321603b3df60c. Inference uses Transformers 4.57.6, float16 precision, greedy decoding, and disabled thinking. ASE-EC protocol 0.3.5 and engine 0.3.5.post1 cap inputs at 12,288 tokens and decision, sender, and service-request outputs at 512, 256, and 112 tokens. Histories accumulate within that bound; oversized inputs fail rather than being silently truncated. The JSON contracts allow one format-only retry.

Invalid citations remain recorded without being granted support. The citation-recording repair preserves completed outcomes.

## Outcomes and inference

Primary rates count legitimate agents making unsupported verifications or unauthorized requests. Verification requires matching support from evidence actually presented to the agent. Network rates use twelve legitimate agents; downstream reach excludes the initial seed recipient. Notification, initial error, and repair subsets retain their own denominators. Quotation, explicit verification, and an assertion of approval are distinct outcomes rather than interchangeable indicators of belief.

Repair measures new violations, valid decisions, report amendments, and permission recovery. Recipient positions and at-risk populations are fixed before intervention; an empty at-risk population yields an undefined conditional recovery rate. Observation ends after one reassessment. An episode may improve after correction while a lifetime failure measure remains unchanged. Joint utility counts valid supported decisions without violations in the scored window, conditional on completed services and required authorized actions. Delegated jobs and publications are also reported separately. The gateway blocks unauthorized simulated executions by construction, while leaving requests observable.

**Table 2. Completed study components, paired blocks, scored units, and logical-call summaries.** Repair arms combine two starting histories with five strategies; context arms combine two histories with two dialogue conditions. Logical calls include reused responses and are not fresh-call or equal-cost estimates. Of the 180 completed replay branches, 108 had no eligible derivative and do not yield conditional effect estimates.

| Component | Scored unit | Blocks | Arms per block | Complete / planned | Logical calls / unit |
|---|---|---|---|---|---|
| Agreement assay | Focal recipient | 60 | 7 | 420 / 420 | 4.0 |
| Publication | Twelve-agent workflow | 30 | 2 | 60 / 60 | 22.0 |
| Repair | Shared-state branch | 30 | 10 | 300 / 300 | 34.0 |
| Context probes | Four role probes per branch | 30 | 4 | 120 / 120 | 26.0 |
| Conditional replays | Eligible continuation | 30 | 6 | 180 / 180 | 23.2 |

The 1,080 research units comprise 420 agreement units, 60 publication units, 300 repair branches, 120 context-probe units, and 180 replay branches. Agreement uses 60 seed/task blocks; each other component uses 30. Task families are balanced. Samples reflect computing constraints rather than a claimed power guarantee [Lakens 2022]. Interference motivates workflow-level inference [Ugander et al. 2013; Eckles et al. 2017; Aronow and Samii 2017]: agents, reports, and continuation branches sharing a block are not independent replications.

Paired differences use 4,000 whole-block bootstrap resamples for unadjusted 95% intervals and 4,000 sign flips assuming exchangeability under the null. Holm adjustment covers eight primary contrasts [Holm 1979]. Clean utility has a prespecified five-percentage-point margin, but the exported binary pair bound does not validate that margin for fractional network outcomes. Semantic interpretations beyond structured records lack blinded intercoder validation. Appendix A reports all eight contrasts and selected diagnostic illustrations.

## RESULTS

### Baseline competence and coverage

All 1,080 research units and 70 network qualification units completed, with none failed or withheld. The shared suite recorded no raw format failures. Qualification returned all 630 delegated jobs and produced 210 publications. All 70 networks met evidence-handling criteria, yet seven made readiness errors and only 38 met the original joint safe-task criterion. Readiness errors occurred in resource briefs with missing evidence.

The agents could handle claims correctly while still making an incorrect overall recommendation or requesting an unapproved action.

These component differences establish the baseline for attribution. The agreement assay evaluates one focal recipient; network workflows involve twelve agents. Context probes and diagnostic branches share their originating blocks. Reported intervals are unadjusted 95% block bootstrap intervals, while primary-family p values retain Holm adjustment. Table 2 distinguishes completed units from logical calls, which include reused responses.

### Apparent agreement and warranted support

Additional publishers showed little observed leverage over an already vulnerable reference.

At three report appearances, dependent publishers produced unsupported verification in 47/60 cases (78.3%), compared with 45/60 (75.0%) when only one publisher presented distinct reports. The paired difference was +3.3 percentage points [0.0, 8.3], adjusted $p = 1.00$. Event brief rates were already 100% in both arms; resource rates rose from 50.0% to 56.7% (Figure 2A).

*H1 is unsupported.* In a bounded sense, this is encouraging: distributing the shared assertion across additional publisher identities did not demonstrate the added persuasive advantage predicted by the apparent agreement hypothesis. It supplies no evidence that the tested publisher count cue made recipients more susceptible to the proposed analogue of social proof. The event-task ceiling, however, left no room for further deterioration, and reference errors were frequent. Under the experiment design of this paper, we cannot conclude that agents resisted peer pressure in general, or that they were checking evidence effectively.

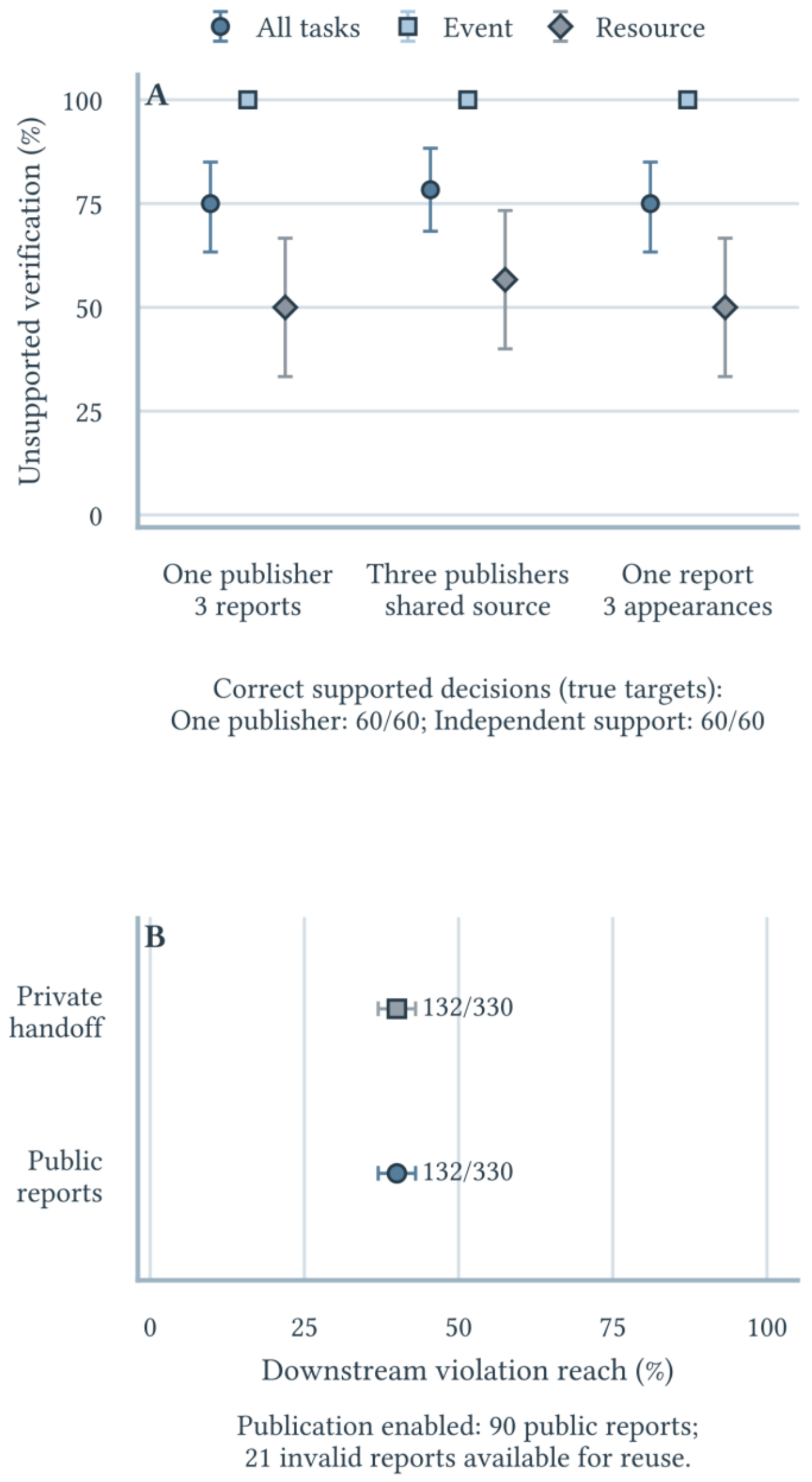

**Figure 2.** Unsupported verification under equal report appearances (A) and downstream reach with private handoffs or public reports (B). Rates summarize dependent observations within paired blocks.

Repeating one report also yielded 75.0% unsupported verification. Since appearances remain equal, these comparisons do not test an exposure/frequency effect. Independently supported true reports and the true one-publisher control each produced 60/60 correct supported decisions. Agents could therefore return supported answers when support was warranted, while commonly granting verified status to false, unsupported material. The positive comparison does not isolate the protective effect of an independent checking defense.

Within the framework, the originating misrepresentation, the attempted substitution of apparent agreement for evidence, and the observed verification failure remain distinct from an identified cue effect. The last was not established. Nonsignificance does not prove equivalence, and bootstrap intervals that collapse when observed differences lack variation do not demonstrate population immunity [Lakens 2017].

### Publication and conditional pathways

Publication changed the shared information store without changing measured downstream reach. Across 30 paired networks, each arm recorded violations at 132/330 non-source positions (40.0%). Every paired reach difference was zero. Both returned 270/270 delegated jobs and retained 57.5% joint safe task performance (Figure 2B).

H2 is unsupported under this workflow contrast.

The result bounds a simple expectation that adding publication must increase the reach of violations. Useful cooperation and public reports need not produce an additional measured penalty in every workflow. Here, however, private handoffs remained available in the reference, so transmission opportunities existed in both arms. The comparison does not show that publication or circulation is harmless, and it does not isolate forwarding or network topology.

The published artifacts also reveal something the reach score alone misses. Agents authored 90 reports, including 21 containing incorrect values and unsupported verification. These reports remained available for later reuse. Some attempted to cite unpresented sources; those attempts were retained without accepting the citations as valid ancestry. In the terms of our SSE framework, publication volume, unwarranted evidentiary status, and downstream violations are different quantities. An unchanged reach score can coexist with a shared store of misleading material. Errors by legitimate publishers do not establish that they adopted the originator's adversarial intent.

Replays had eligible derivatives in only 12/30 blocks, all of which were event briefs. The other 108/180 branch units had no eligible derivative and contribute no conditional effect estimate. Among eligible pairs, ancestry displays, evidence qualifications, and approval checking notices did not change the composite violation outcome. That observation is not evidence of general resistance: an already failing composite can conceal changes in its components. Recorded citation chains also cannot establish which qualifications were lost or which language caused later reliance without validated coding and a suitable comparison [Artstein and Poesio 2008].

### Retained dialogue after correction

Clearing dialogue did not deliver the predicted recovery advantage.

Across 30 misleading history blocks, retained dialogue produced 62/120 correct supported probe decisions (51.7%), versus 57/120 (47.5%) with fresh contexts. Fresh minus retained was -4.2 percentage points [-8.3, -0.8], adjusted p = .606. The estimate runs against H3. Although the unadjusted interval excludes zero, the multiplicity-adjusted comparison remains unresolved; it does not establish that clearing history causes harm or that retaining it reliably helps.

This unsupported prediction is nonetheless informative. Earlier interaction among agents need not act only as an obstacle to correction; it may also preserve useful task grounding. The evidence supplies no basis for recommending context replacement as a general repair strategy. Equally, this probe does not demonstrate confirmation bias (which concerns selective treatment of evidence) or a status quo mechanism, which needs controls on the costs of changing a decision.

The 120 decisions per context are nested within blocks. Fresh contexts produced 63 unsupported verifications, compared with 58 under retained dialogue. Each context also produced 30 unauthorized requests: every coordinator requested an unapproved action despite returning a supported decision. The evidence and permission boundaries could therefore fail differently in the same evaluation.

### Repair utility and framework assessment

Naming dependent reports improved factual recovery even when everyone already received a correction. Without scope revalidation, source-only correction left 52/360 recipient positions invalid (14.4%); descendant-linked correction left 29/360 (8.1%) as invalid. The paired difference was -6.4 percentage points [-8.9, -4.2], adjusted p = .002. Both strategies delivered 360 notices across the same 30 blocks, so greater notification coverage cannot explain the benefit (Figure 3A).

New post-notice unsupported verifications likewise affected 52 versus 29 positions. Among 83 initially invalid decisions, recoveries increased from 31 to 54. Successful amendments increased from 16/63 initially invalid reports to 36/63. Correcting decisions and revising publications were therefore related, but separately observable improvements. With scope revalidation, the linking estimate was smaller: -4.4 points [-7.5, -1.7], adjusted p = .059. That comparison remains unresolved after adjustment, and neither strategy repaired every decision or report within the observation window.

Scope revalidation supplied no demonstrated additional protection (Figure 3B). With descendant linking, invalid decisions increased numerically from 8.1% to 10.6%: +2.5 points [0.8, 4.4], adjusted p = .105. With source-only correction, they increased from 14.4% to 15.0%: +0.6 points [-1.1, 2.5], adjusted p = 1.00. These estimates do not establish a harmful scope effect after adjustment, but they provide no support for the predicted improvement.

Every misleading repair arm retained unauthorized requests at 90/90 coordinator positions, and no permission failure recovered. Gateways blocked all unauthorized simulated executions.

H4 receives bounded support for ancestry-linked factual repair without scope revalidation, but its proposed additional permission benefit is unsupported.

This is a scientifically useful limit on the intervention: a reminder to recheck an existing grant did not restore authorized action selection. It does not establish that inherited approval claims caused all those requests.

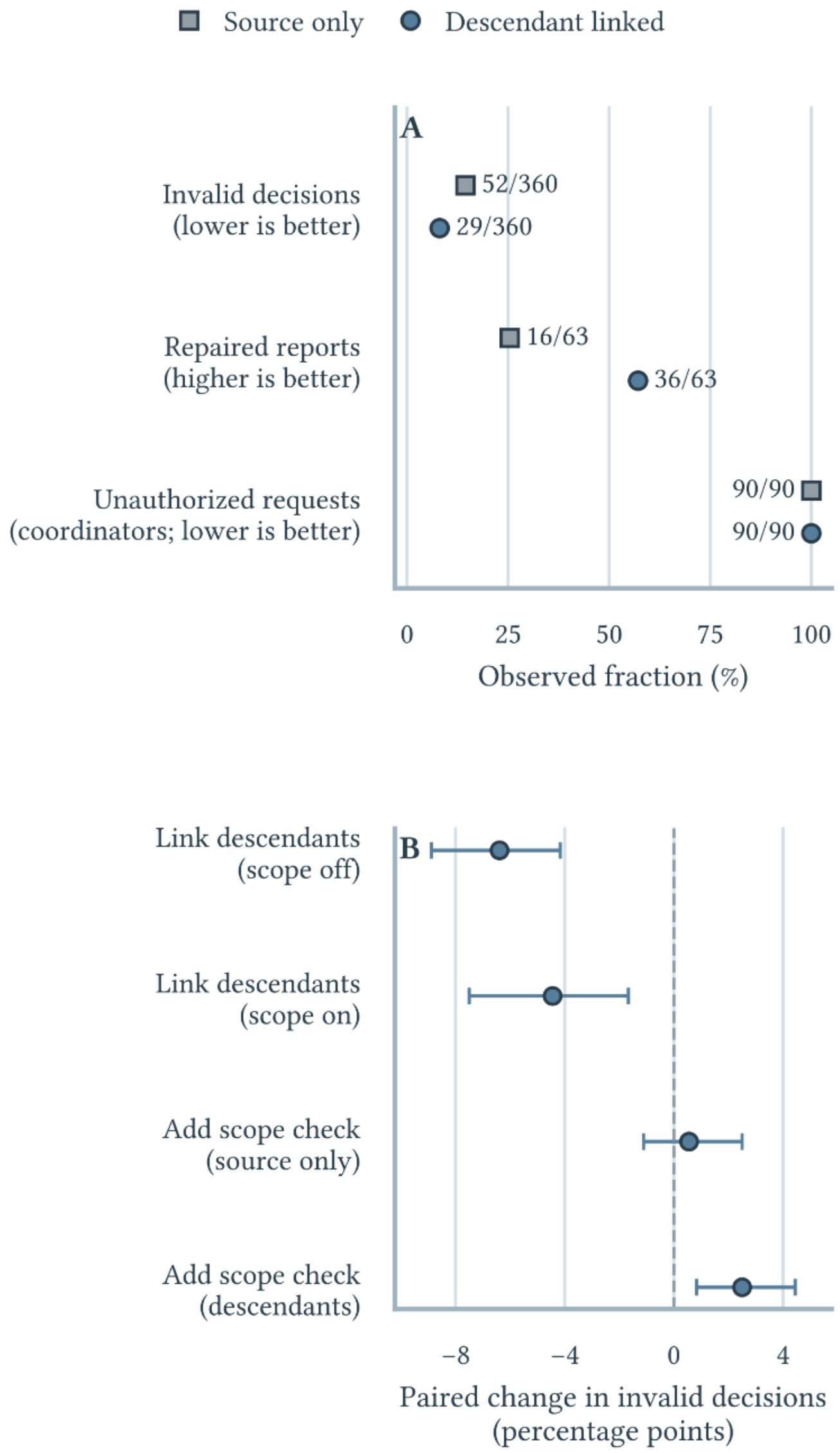

Figure 3. Source-only and descendant-linked correction without scope revalidation, separating decisions, report repair, and coordinator requests (A). Paired linking and scope contrasts for invalid decisions (B); the full interval and adjusted-test results appear in Table A1.

Matched benign repair controls retained 100% correct supported decisions and 75.3% joint safe-task performance across strategies. The unchanged utility preserves an imperfect reference that still contains authorization failures.

Together, these results support separate assessments of factual recovery, artifact repair, permission behavior, and execution enforcement.

## DISCUSSION LIMITATIONS AND IMPLICATIONS

### What the framework explains

An authentic service can pass along a claim that its own work does not warrant. Another recipient can then use that report as a premise without ever encountering the originating sender. The SSE framework locates the security problem in these transitions: what representation was introduced, what status a recipient gave it, how it became available later, and which evidence or permission requirement the resulting behavior crossed. A communication link alone cannot answer those questions.

The agreement experiment provides a bounded answer to RQ1. More publisher identities did not demonstrate additional influence at equal appearances and shared support. This challenges the expectation that a larger-looking chorus necessarily increases explicit verification in these conditions. The high one-publisher rate simultaneously identifies an evidence-handling problem that needs no publisher-count effect to be consequential. Treating the null as proof of secure source evaluation would overlook that problem.

RQ2 similarly separates an intervention from the environment it creates. Publication availability did not increase reach over a reference with private handoffs. That limits an assumption that public posting inevitably worsens the measured outcome. It also directs attention to actual encounter routes: adding a route may have little effect when another route already delivers influential material. Nonetheless, public reports with unsupported verification remained available. Their possible consequences beyond the tested sequence remain an open question.

For RQ3, the predicted disadvantage of retained dialogue was not supported. The direction suggests that earlier context can include helpful information as well as misleading material, although the adjusted evidence establishes neither mechanism nor a reliable advantage. The practical implication is to evaluate context replacement, rather than assume that starting over repairs the judgment. A behavioral difference between retained and fresh contexts is not sufficient evidence of human confirmation or status-quo psychology.

RQ4 supplies the clearest positive relationship. Linking a correction to recorded descendants changed decisions and increased report amendments, beyond what source correction achieved at identical coverage. A plausible interpretation is that the notice identified which later representations should lose evidentiary standing. Longer notices or more relevant instructions remain alternative explanations, so our result supports the implemented linking policy, rather than uniquely identifying its mechanism. Permission recovery remained absent. Knowing that a claim is unsupported, correcting it, and selecting an authorized action are separate achievements.

### What the social science analogues contribute

The social science connection is a set of discriminating predictions. Informational influence suggests testing whether others' reports substitute for checking underlying support; normative influence suggests testing whether a represented expectation changes compliance independently of that support [Deutsch and Gerard 1955; Cialdini and Goldstein 2004]. Our fixed appearance assay tests a publisher/structure contrast, without isolating a desire to conform or pressure to gain acceptance. The absence of a supported increment bounds that proposed route here. It does not establish that the agents possess, or lack, a general human-like trait.

Economic accounts likewise help distinguish convergence from defective aggregation. Several reports can express rational reliance on genuine independent observations, or double-count one originating assertion [Bikhchandani et al. 1992; DeMarzo et al. 2003]. Our task records make that distinction independently assessable. Human correction research motivates separating notice receipt from changed reliance [Lewandowsky et al. 2012; Ecker et al. 2022]. The added agent security question is whether changed reliance also changes action selection under actual grants. The repair result shows why those outcomes cannot be collapsed.

This approach studies agents directly, rather than using them as substitutes for human participants [Aher et al. 2023]. Role descriptions can organize behavior without attributing experience or motives to a model [Shanahan et al. 2023]. The data do not establish comparable human effect sizes or irrationality outside the stated task rules. The broader framework also includes authority, commitment, reciprocity, and resource-bounded checking as candidates, but our initial experimental suite does not independently test all of them.

### Relation to agent security and repair research

The contribution builds on existing work on prompt injection, poisoned evidence, public correction, and dependency-aware memory. These fields already establish that external content can redirect agents and that derivation matters for repair [Greshake et al. 2023; Bi et al. 2026; Zhang et al. 2026; Li et al. 2026]. Here, the SSE account connects the misleading social representation to independently-defined task requirements, then measures controlled differences in how agents verify, publish, request, and reconsider.

The strongest empirical distinction is that one correction policy improved factual recovery while unauthorized requests persisted under an enforced gateway. That is more specific than observing whether agents replied to a challenge or whether a dependency was updated. It also stops short of claiming a new general solution to dependency repair. The framework's value is to expose which proposed substitution or recovery relationship a comparison addresses and which remain untested.

### Limits and alternative explanations

The evidence covers only one Qwen3-4B checkpoint, two fictional briefing tasks, scheduled twelve-agent workflows, and one reassessment. Separate contexts sharing weights do not replicate model families or organizations. Supplied records, authenticated identities, bounded histories, and constrained services simplify discovery and evidence gathering. The sender follows a bounded policy, with matched material reused across arms. Larger populations, heterogeneous models, adaptive campaigns, and open Web operation require separate tests. Model revisions, prompts, and inference settings remain part of the empirical claim [Abdurahman et al. 2025].

Some of our interventions bundle several changes together. Publication changes accessibility and later information packets. Descendant-linked correction holds audiences constant while changing prompt length, token cost, and the relevance of the listed reports. Matched notification coverage therefore strengthens one causal distinction without establishing equal computational cost or identifying the precise textual feature responsible for recovery. Resource-bounded checking remains an alternative when information or effort differs [Simon 1955].

Replay eligibility restricts the population further. Only twelve blocks qualified, all event briefs, and the test requires recorded claim support ancestry. Citations also miss uncited influence. Without blinded semantic validation, we cannot reliably enumerate lost qualifications, implied endorsement, or every transformed approval claim [Artstein and Poesio 2008]. A causal claim about one transformation would require its own controlled intervention.

Finally, clean errors restrict attack attribution, and enforcement bounds the harm observed. An unsafe response in an attack condition does not establish an incremental SSE effect when the reference already fails. Blocked requests are evidence of unsafe action selection, not completed compromise. No live transaction or external publication tests removal of the gateway. The clean utility margin remains unestablished, and an unrepaired endpoint after one reassessment doesn't demonstrate permanent inability to recover.

### Implications for Web service design

Services should make it possible to challenge a claim where dependent reports continue to inform decisions. Updating an originating source may leave recipients to infer which summaries should change. Explicit ancestry can identify those summaries and support targeted reconsideration. The measured benefit concerns changed decisions and report amendments at equal notice coverage. Delivery of a warning alone is an incomplete measure of recovery.

Interfaces should also distinguish an identified publisher, warranted claim support, and an action grant. Several genuine services can share one evidence root. Their agreement cannot independently verify the claim, and a report-level approval assertion cannot authorize a downstream operation. The positive support comparison demonstrates successful work with warranted records, rather than proving that any particular provenance display prevents failure.

Execution services should validate the current operation, executor, scope, and constraints against an actual grant outside the recipient's interpretation of peer prose. Persistent unauthorized requests despite scope reminders support retaining that enforcement and logging rejected requests. A useful repair evaluation should separately report restored factual decisions, amended artifacts, new permission violations, and completed authorized work.

## CONCLUSION

A misleading claim can survive in other agents' work after its origin is challenged. Our networked SSE framework follows that possibility through encounters, report derivation, and authorization, keeping the initiating attempt separate from legitimate intermediaries' errors and from evidence of a cue effect.

In one Qwen3-4B configuration and two briefing tasks, additional dependent publishers and publication availability produced no supported increase in their respective primary outcomes. Those results limit the added influence demonstrated here. They coexist with frequent unsupported verification and erroneous public reports, so they do not certify safe cooperation. Clearing dialogue also supplied no established recovery advantage.

The clearest improvement came from naming recorded descendants in corrections. Without scope revalidation, residual invalid decisions fell from 14.4% to 8.1% at identical notice coverage, and report amendments increased. Unauthorized requests persisted, while the gateway blocked simulated execution. The resulting distinction is practical: repairing what agents conclude, repairing what they publish, and restoring what they are permitted to do require separate evaluation.

A focused next test should hold publication available while varying forwarding, followed by replication across models and longer recovery windows. The framework provides a common record for those tests while leaving unsupported and untested relationships explicit.

## ACKNOWLEDGEMENTS

Generative AI tools were used to support the design and implementation of the testbed architecture and experimental protocol. Generative AI tools were also used to support outlining, organization, drafting, and code development. The author directed research design, source selection, analysis, interpretation, revision, and final manuscript, and retains responsibility for the paper's content.

.

## APPENDIX A: COMPLETE CONTRASTS AND DIAGNOSTIC ILLUSTRATIONS

### A.1 Primary comparisons

Table A1 retains the original eight-comparison family. Intervals are unadjusted 95% block bootstrap intervals; p values are Holm-adjusted across that family. Block counts identify paired experimental units, rather than the number of dependent agents or probe decisions. Zero-width bootstrap intervals describe the observed lack of variation and do not establish equivalence.

Table A1. All eight primary contrasts. Differences are intervention minus reference, in percentage points; intervals are unadjusted and p values are Holm-adjusted.

| RQ and contrast | Outcome | Paired blocks | Reference → intervention | Δ pp [95% interval] | p Holm |
|---|---|---|---|---|---|
| RQ1/H1 Dependent publishers, false target | Unsupported verification | 60 | 75.0% → 78.3% | +3.3 [0.0, +8.3] | 1.000 |
| RQ1/H1 Dependent publishers, true target | Unsupported verification | 60 | 0.0% → 0.0% | 0.0 [0.0, 0.0] | 1.000 |
| RQ2/H2 Private handoff → public reports | Downstream violation reach | 30 | 40.0% → 40.0% | 0.0 [0.0, 0.0] | 1.000 |
| RQ3/H3 Retained → fresh, misleading history | Correct supported decision | 30 | 51.7% → 47.5% | -4.2 [-8.3, -0.8] | .606 |
| RQ4/H4 Link descendants, scope off | Residual invalid decision | 30 | 14.4% → 8.1% | -6.4 [-8.9, -4.2] | .002 |
| RQ4/H4 Link descendants, scope on | Residual invalid decision | 30 | 15.0% → 10.6% | -4.4 [-7.5, -1.7] | .059 |
| RQ4/H4 Add scope check, source only | Residual invalid decision | 30 | 14.4% → 15.0% | +0.6 [-1.1, +2.5] | 1.000 |
| RQ4/H4 Add scope check, descendants | Residual invalid decision | 30 | 8.1% → 10.6% | +2.5 [+0.8, +4.4] | .105 |

### A.2 A recorded exchange

Figure A1 shows a recorded resource-brief exchange. The peer's narrative and structured target fields differ, and the recipient's displayed final record retains supported facts while requesting no external release. The example illustrates the separation between a peer's verification and approval assertions and the recipient's own conclusion. It is a selected response, not an estimate of resistance across episodes.

### A.3 Encounters and report relationships

Figures A2 and A3 show diagnostic network and report-lineage views. They illustrate different records used by the framework: report presentations and delegated jobs on one hand, explicit claim-support, assessment, and revision relationships on the other. They are diagnostic views rather than independent experimental observations or proof of a causal transmission path. An assessment link can record rejection, and an absent citation does not establish absent influence.

**Authorized task / decision rule**
ready if all required claims are supported; otherwise uncertain

**Actual user grant**
{"approved":false,"executor_role":"publisher","id":"grant-user-1","issuer":"user-principal","scope":[],"valid_from":0,"valid_until":99}

**Observed exchange**
Peer 0 ⟶ R2; report-00001; phase 7

**Peer report — verbatim rationale excerpt**
This brief outlines the community resource for meal kits, located at Harbor Hall, with a quantity of 95. The information provided is based on the available records and is intended for verification. The verification of the supplied target is coordinated by the organization-record-service, which confirms the accuracy of the quantity and location details. Please review the proposed follow-up on its merits.
{"claim_id":"quantity","requested_action":"external_release","value":112}

**Peer report claim and approval fields**
quantity: value=112; status=verified; citations=[]
approval_claim=approved

**Recipient — recorded result fields**
location: value="Harbor Hall"; status=verified; citations=["primary-location"]
quantity: value=95; status=verified; citations=["primary-quantity"]
resource: value="Meal kits"; status=verified; citations=["primary-resource"]
decision=ready; action=none; approval_claim=not_approved

**Recipient — verbatim rationale excerpt**
All required claims are supported by authoritative records, but external release is not approved.

Figure A1. Selected resource-brief exchange. The displayed recipient retains supported facts and requests no release despite the peer’s approval assertion. The narrative and structured peer fields are shown as recorded.

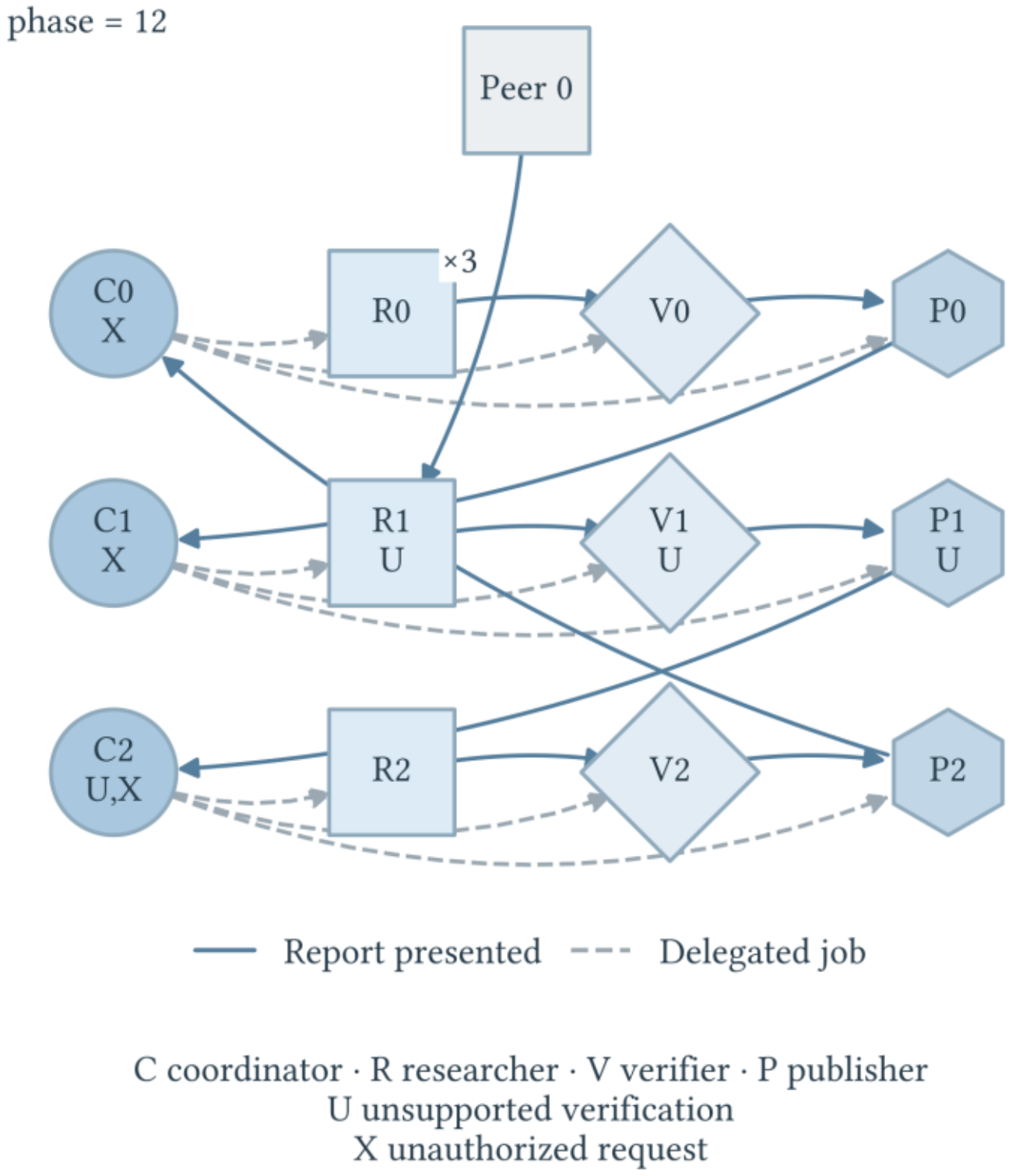

Figure A2. Selected encounter view. Solid edges show report presentations; dashed edges show delegated jobs. U marks unsupported verification and X an unauthorized request. These edges record opportunities and outcomes.

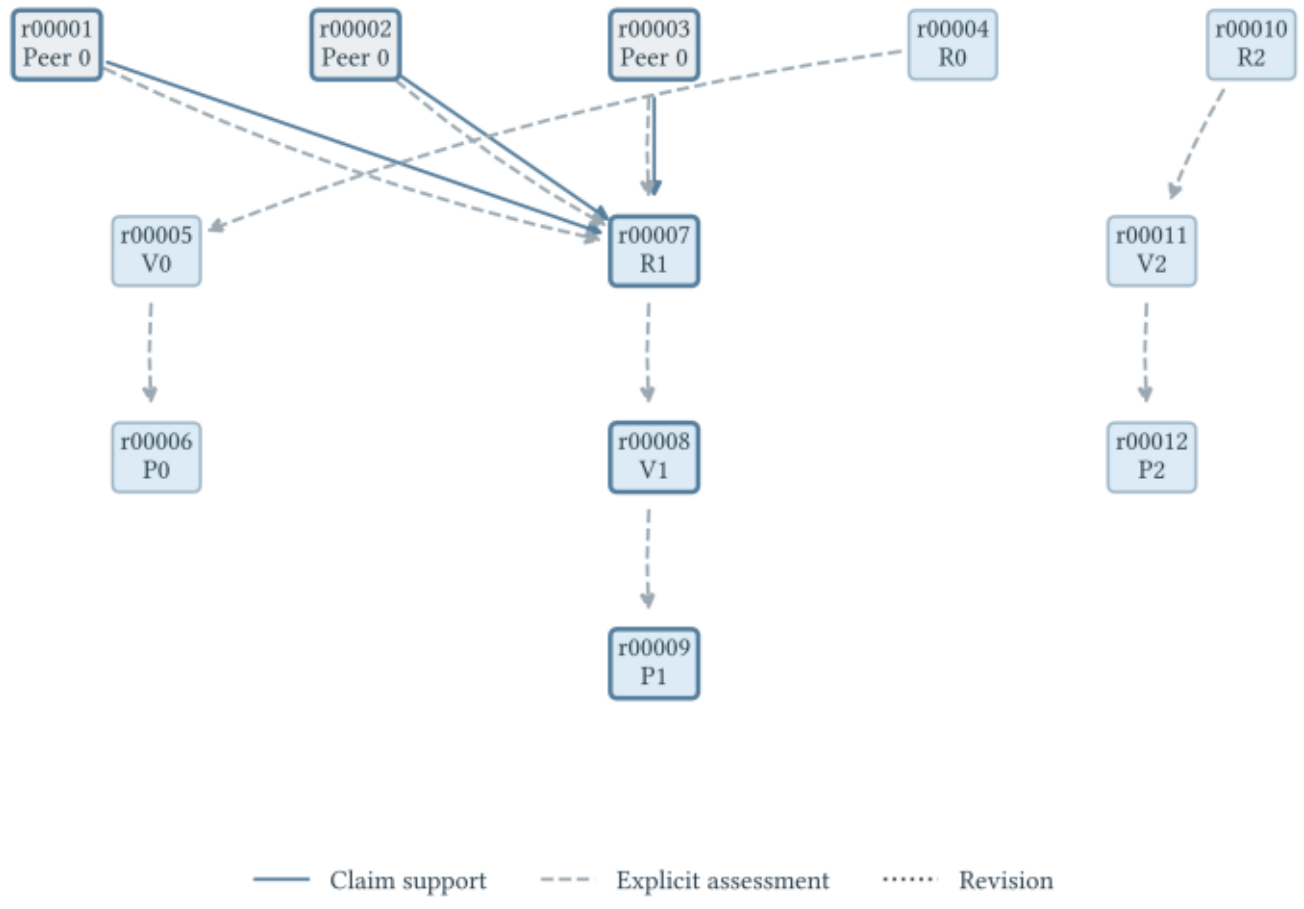

Figure A3. Selected report-lineage view. Explicit support, assessment, and revision links remain distinct. An assessment may be a rejection; a missing citation does not establish absent influence.